\documentstyle[aps,epsf,amssymb]{revtex}  

\begin{document}        

\baselineskip 14pt
\title{Neutrino-induced upward-going muons in Super-Kamiokande}
\author{A.~Habig, for the Super-Kamiokande Collaboration}
\address{Boston University}
%
\maketitle              

\begin{abstract}        
  Upward-going muons observed by the Super-Kamiokande detector are
  produced by high-energy atmospheric neutrinos which interact in rock
  around the detector.  Those which pass completely through the detector
  have a mean parent neutrino energy of $\sim 100$~GeV, while those
  which range out inside the detector come from neutrinos of mean energy
  $\sim 10$~GeV.  The neutrino baseline varies with the observed muon
  zenith angle, allowing for an independent test via $\nu_\mu$
  disappearance of the neutrino oscillations observed in the
  Super-Kamiokande contained events.  614 upward through-going and 137
  upward stopping muons were observed over 537 (516) live days,
  resulting in a flux of $\Phi_t=1.74\pm0.07 {\rm(stat.)}\pm0.02{\rm
    (sys.)}$ ($\Phi_s=0.380\pm0.038 {\rm(stat.)}^{+0.019}_{-0.016}{\rm
    (sys.)}$) $\times10^{-13} {\rm{cm^{-2}s^{-1}sr^{-1}}}$.  The
  observed stopping/through-going ratio ${\mathcal{R}}=0.218\pm
  0.023{\rm(stat.)}^{+0.014}_{-0.013}{\rm(syst.)}$ is $2.9\sigma$ lower
  than the expectation of $0.368^{+0.049}_{-0.044}{\rm(theo.)}$.  Both
  the shape of the zenith angle distribution of the observed flux and
  this low ratio are inconsistent with the null oscillation hypothesis,
  but are compatible with the previously observed
  $\nu_\mu\leftrightarrow\nu_\tau$ oscillations.  Taken as a whole, the
  addition of these higher energy $\nu_\mu$ data to the contained
  neutrino events provides a better measurement of the oscillation
  parameters, narrowing the allowed parameter range to $\sin^22\theta
  \gtrsim 0.9$ and $1.5\times10^{-3}eV^2 \lesssim \Delta m^2\lesssim
  6\times10^{-3}$ at 90\% confidence.
\end{abstract}          

\section{Introduction}               

The Super-Kamiokande (``Super-K'') atmospheric neutrino analyses
presented in\cite{skosc} and updated in these
proceedings\cite{messier-dpf} provide evidence for
$\nu_\mu\leftrightarrow\nu_\tau$ flavor oscillations.  These analyses
examine neutrinos whose interaction vertices are inside the water of the
Super-K detector\cite{sksubgev,skmultigev}.  However, to extend this
analysis to higher energies, where the steeply falling power law
spectrum of the neutrinos' cosmic ray parents reduces the absolute flux
of atmospheric neutrinos, a larger interaction volume must be used.
This is accomplished by studying the products of neutrino interactions
in the rock surrounding the detector.

Energetic atmospheric $\nu_{\mu}$ or $\bar{\nu}_{\mu}$ interact
with the rock surrounding the Super-K detector and produce muons via
weak interactions.  While the constant rain of cosmic ray muons
overwhelms any downward-going neutrino induced muons, upward-going muons
are neutrino induced because upward-going cosmic ray muons cannot
penetrate the whole Earth.  The flavor of these parent neutrinos is
$\nu_\mu$ or $\bar{\nu}_\mu$, as $\nu_e$ and $\bar{\nu}_e$ induced
electrons and positrons shower and die out in the rock before reaching
the detector.  Detection of upward-going muons resulting from from tauons
produced in $\nu_\tau$ or $\bar{\nu}_\tau$ charged current interactions
is suppressed by branching ratios and kinematics to $\lesssim 3\%$ of
the $\nu_\mu$ induced muon flux.  Thus, the upward-going muon technique
results in a reasonably pure sample of muon-flavored neutrinos, allowing
a test of possible $\nu_\mu$ disappearance due to flavor oscillations.

\begin{figure}[thbp]
  \centerline{\epsfxsize 4.0 truein \epsfbox{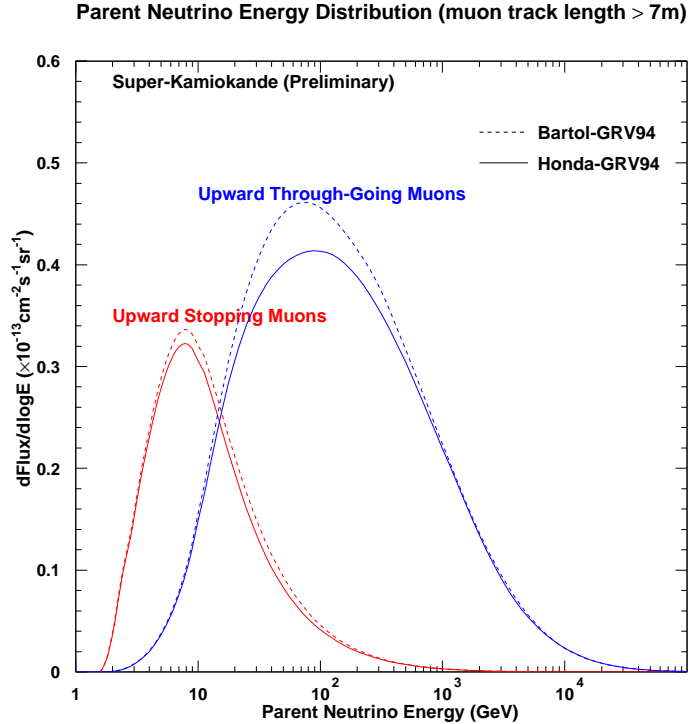}}   
  \caption{The energy spectra of the parent neutrinos of stopping
    upward-going muons (left) and upward through-going muons (right).
    The dashed lines are the result of using the Bartol input fluxes, and
    the solid lines are from those of Honda.}
  \label{fig:spectra}
\end{figure}

Those muons energetic enough to cross the entire detector are defined as
``upward through-going muons''\cite{skupthru}.  The mean energy of their
parent neutrinos is approximately 100~GeV.  Those upward-going muons
that range out inside the detector are defined as ``stopping
upward-going muons'', and come from parent neutrinos with a mean energy
of about 10~GeV.  In comparison, the ``Partially Contained'' events
from\cite{skmultigev} have a similar parent neutrino spectrum to that of
the stopping upward-going muons, the ``Multi-GeV'' neutrinos
from\cite{skmultigev} typically have several GeV of energy, and the
``Sub-GeV'' neutrinos from\cite{sksubgev} are less than a GeV.  The
upward-going muon parent neutrino spectra are shown in
Fig.~\ref{fig:spectra}.

\section{The Experiment}

The Super-K detector is a 50~kton cylindrical water Cherenkov
calorimeter located at the Kamioka Observatory and administered by the
Institute for Cosmic Ray Research of the University of Tokyo.  To reduce
the cosmic ray muon background, the detector is placed $\sim$1000~m
underground in the Kamioka mine, Gifu prefecture, Japan.  The detector
is divided by an optical barrier instrumented with photomultiplier tubes
(``PMT''s) into a cylindrical primary detector region (the Inner
Detector, or ``ID'') and a surrounding shell of water (the Outer
Detector, or ``OD'') which allows the tagging of entering and exiting
particles.  Details of the detector can be found in
Ref.~\cite{sksubgev}.  An upward-going muon is defined as an event that
appears to enter from the rock and is traveling upwards.  Thus, PMT
activity in the OD at the muon's entrance point is required, and those
events which also have an OD signal at the muon's exit point are
classified as through-going.  Note that a neutrino interaction inside
the water of the OD itself also produces an ``entering'' signal, so some
small fraction ($\lesssim 1.5\%$) of the upward-going muon sample
actually originates in the OD rather than the rock.  This effect is
accounted for in the expected flux calculations.

The cosmic ray muon rate at Super-K is 2.2~Hz.  The trigger efficiency
for a muon entering the detector with momentum more than 200~MeV/$c$ is
$\sim$100\% for all zenith angles.  The nominal detector effective area
for upward-going muons with a track length \(>\) 7m in the ID is
$\sim$1200~m$^2$.  

The data used in this analysis were taken from Apr.~1996 to Jan.~1998,
corresponding to 537 days of detector livetime for the through-going
muon analysis and 516 live-days of the stopping muon analysis.  Event
reconstruction is made by means of the charge and timing information
recorded by each hit PMT.  The direction of a muon track is first
reconstructed by several automated grid search methods, which find the
track by minimizing the width of the residual distribution of the photon
time-of-flight subtracted ID PMT times.  Finally, an independent double
hand-scan of all upward-going muon candidates (event loss probability
\(<\) 0.01\%) is then done to eliminate bad fits and to obtain a final
precision fit.

A minimum track length cut of 7m ($\sim$1.6~GeV) was applied.  This cut
serves to eliminate short pathlength events that are very close to the
PMTs and thus hard to reconstruct, as well as providing a large
energy threshold and eliminating non-muon showering background.  
To reduce the abundant downward-going cosmic ray muons, events
satisfying $\cos\Theta<0.1$ are selected, where $\Theta$ is the zenith
angle of the muon track, with $\cos\Theta<0$ corresponding to
upward-going events, and the down-going muons satisfying
$0<\cos\Theta<0.1$ are used to estimate the up/down separation resolution
near the horizon.  After a visual scan by two independent groups and a
final direction hand-fit, 614 through-going and 137 stopping
upward-going muon events with $\cos\Theta<0$ remain.  

Due to the finite angular resolution ($1.5^\circ$) and multiple Coulomb
scattering in the nearby rock, some down-going cosmic ray muons may
appear to have $\cos\Theta<0$.  The estimation of this background in the
nearly horizontal zenith angle bin is done by fitting the nearly
horizontal down-going cosmic ray muon zenith angle shape and projecting
its tail below the horizon (see \cite{skupthru} for a detailed
discussion).  This background is estimated to be 4.3$\pm0.4$
through-going and 13.2$\pm3.5$ stopping events, all contained in the
last (closest to horizontal) zenith angle bin.  The stopping muon sample
has a larger contamination than the through-going sample due to the
lower energies allowed in the stopping muon sample, and is more
uncertain due to lower statistics.  The contamination at the Kamioka
site due to cosmic ray photoproduced upward-going pions\cite{up-pions}
and electromagnetic showers (from $\nu_e$ charged current or neutral
current interactions) meeting the 7m track length requirement is
estimated to be \(<\) 1\%.

The total detection efficiency of the complete data reduction process
for upward through-going muons is estimated by a Monte Carlo simulation
to be \(>\)99\% which is uniform for $-1 < \cos\Theta < 0$.  Using the
upward/downward symmetry of the detector configuration, the validity of
this Monte Carlo program has been confirmed by downward-going cosmic ray
muon data.

\begin{figure}[thbp]
  \centerline{
    \begin{minipage}[t]{3.5in}
      \epsfxsize 3.5 truein \epsfbox{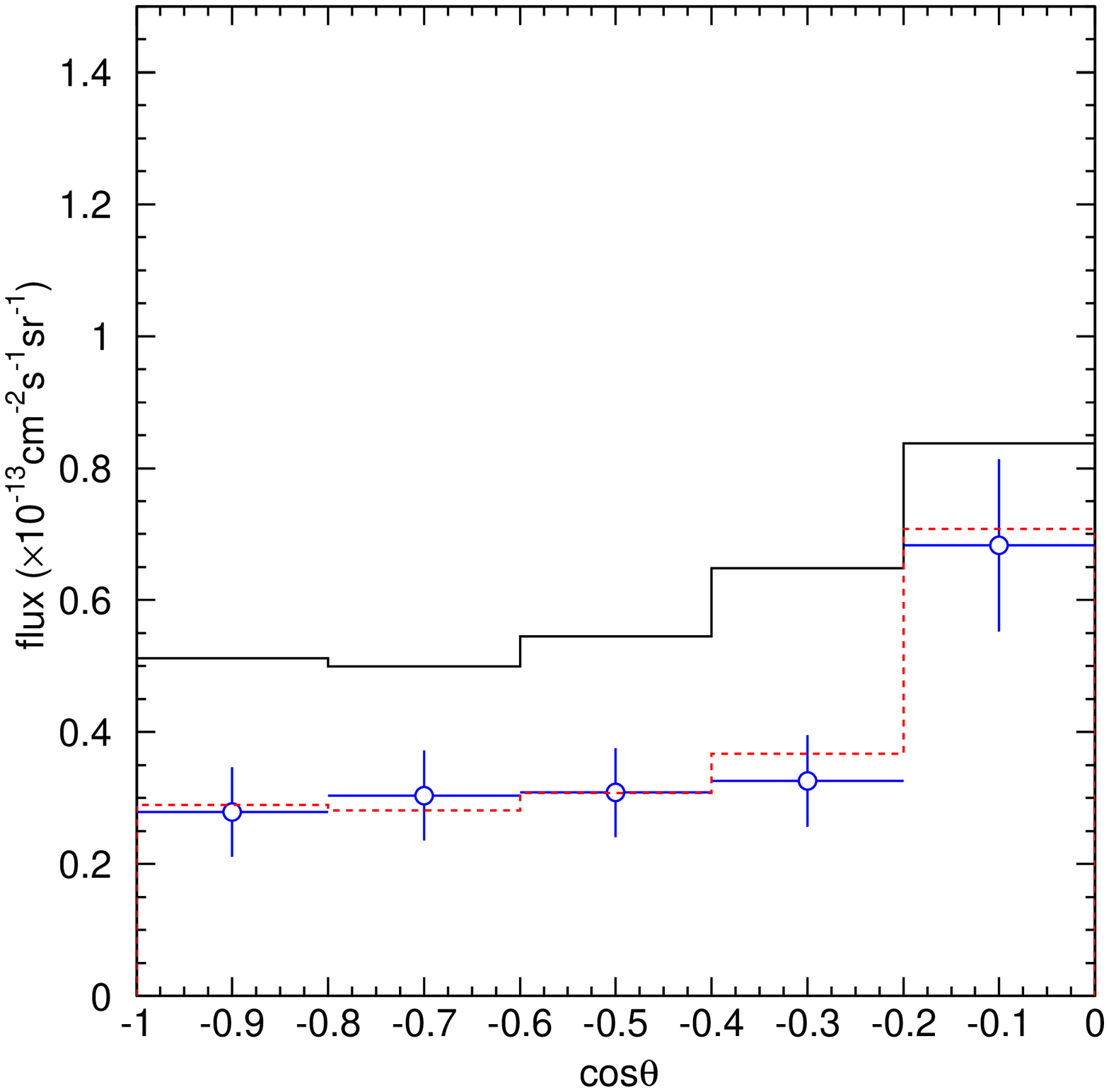}
    \end{minipage}
    \begin{minipage}[t]{3.5in}
      \epsfxsize 3.5 truein \epsfbox{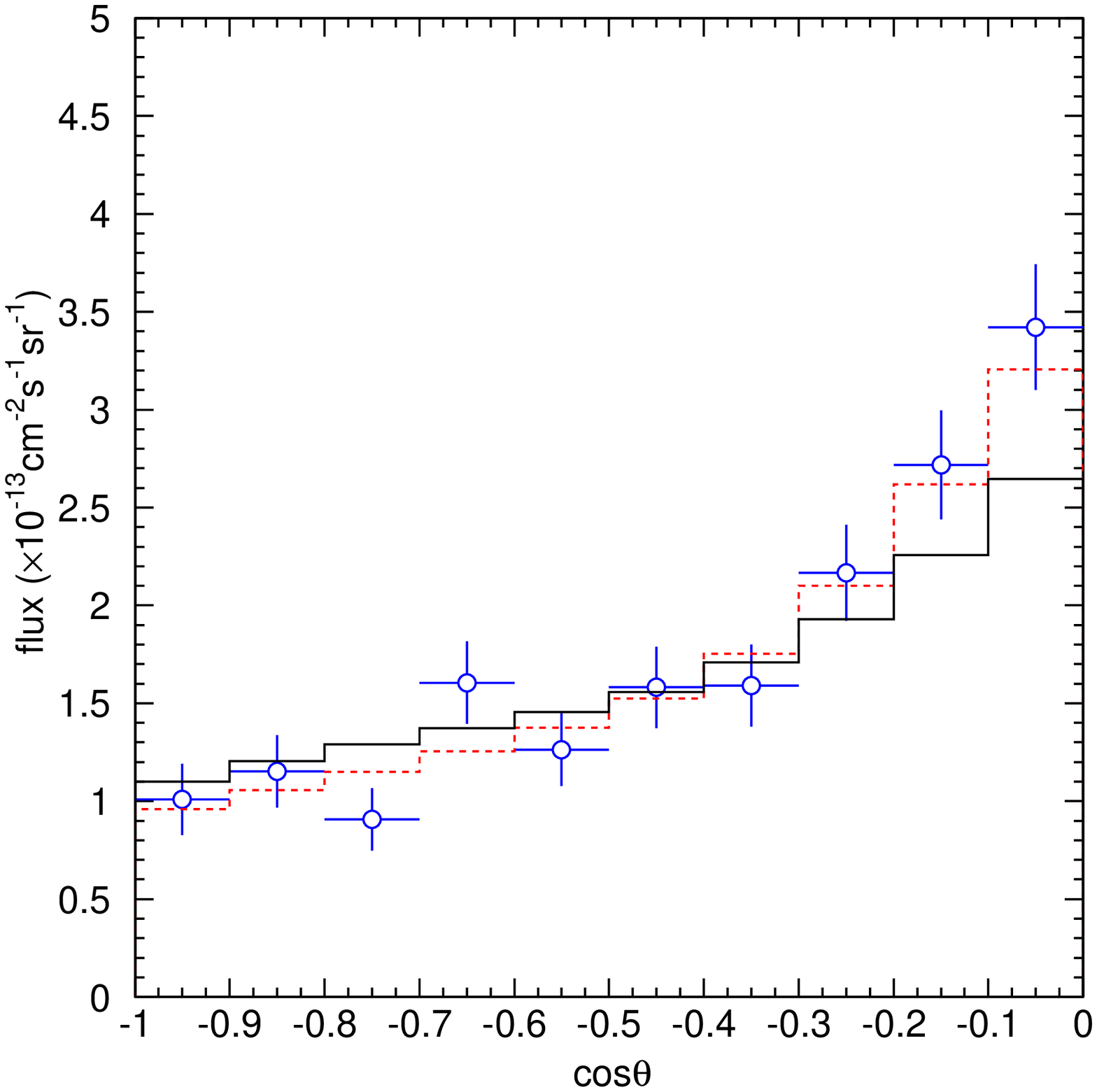}
    \end{minipage}
    }   
  \caption{Stopping (a) and through-going (b) upward-going muon 
    fluxes observed in Super-K as a function of zenith angle.  The error
    bars indicate uncorrelated experimental systematic plus statistical
    errors added in quadrature.  The solid histogram shows the expected
    fluxes based on the Honda/GRV94 model for
    the null neutrino oscillation case, normalized by $\alpha=-10\%$
    from the best-fit no-oscillation through-going muon case.  The
    dashed lines are the expected flux in the presence of
    $\nu_\mu\leftrightarrow\nu_\tau$ oscillations at maximal mixing and
    $\Delta m^2=3.2 \times 10^{-3}$eV$^2$, the best-fitting oscillation
    hypothesis.}
  \label{fig:fluxzen}
\end{figure}

The resulting total flux observed in each of the two upward-going muon
samples integrated over the whole lower hemisphere is listed in
Table~\ref{tab:results}.  The flux plotted as a function of zenith angle 
can be seen in Fig.~\ref{fig:fluxzen}.  The lower statistics available
for the stopping upward-going muon analysis is the reason for the coarser 
binning of those events.

\section{Expected Flux}

To predict the expected upward-going muon flux, this analysis used a
model which is a combination of the atmospheric neutrino flux from Honda
{\it et al}~\cite{honda} and a neutrino interaction model composed of
quasi-elastic scattering~\cite{qe} + single-pion production~\cite{sp} +
deep inelastic scattering (DIS) multi-pion production.  The DIS
cross-section is based on the parton distribution functions (PDF) of
GRV94DIS~\cite{grv94} with the kinematic constraint of
$W>1.4$~GeV/c$^2$.  Lohmann's muon energy loss formula in standard
rock~\cite{lohmann} is then employed to analytically calculate the
expected muon flux at the detector.  This flux is compared to three
other analytic calculations to estimate the model-dependent
uncertainties of the expected muon flux.  The other flux calculations
use the various pairs of the Honda flux, the GRV94DIS PDF, the
atmospheric neutrino flux model calculated at Bartol~\cite{bartol},
and the CTEQ3M~\cite{cteq} PDF.  These comparisons yield the theoretical
systematic errors used in the calculations.  In this analysis, Honda's
input flux is used as the primary input so as to be more directly
comparable with the contained event analysis\cite{skosc}, which is also
based upon the Honda flux.

The Honda+GRV94DIS calculation results in an expected muon flux shown
in Table~\ref{tab:results} for $\cos\Theta < 0$.  The dominant error in
the absolute flux comes from the absolute normalization uncertainty in
the input neutrino flux, which is estimated to be approximately
$\pm20$\%~\cite{bartol,honda,frati} for neutrino energies above several
GeV.  The dominant error in the ratio comes from the uncertainty in the
slope of the cosmic ray spectrum.  The parent neutrino energy spectrum
is shown in Fig.~\ref{fig:spectra}.  Note the clear separation in
energies of the stopping and through-going upward-going muon parents.

\begin{center}
  \begin{minipage}[t]{5.0in} 
    \begin{table}
      \caption{Observed and expected (Honda-GRV94) 
        upward-going muon fluxes, in units of 
        $10^{-13} {\rm{cm^{-2}s^{-1}sr^{-1}}}$, and their ratio
        ${\mathcal{R}}=\Phi_s/\Phi_t$.}
      \begin{tabular}{l|c|c|c|cc} 
        & $N_{obs}$ & $N_{exp}$ & Observed $\Phi$ & Expected $\Phi$\\\hline
        Through & 614 & 695
                & $1.74\pm0.07 {\rm(stat.)}\pm0.02{\rm (sys.)}$
                & $1.84\pm0.40(\rm{theo.}) $\\  
        Stop    & 137 & 244 
                & $0.380\pm0.038 {\rm(stat.)}^{+0.019}_{-0.016}{\rm (sys.)}$
                & $0.676\pm0.149(\rm{theo.}) $\\\hline  
        \multicolumn{2}{l}{${\mathcal{R}}$} &
                & $0.218\pm 0.023{\rm(stat.)}^{+0.014}_{-0.013}{\rm(syst.)}$
                & $0.368^{+0.049}_{-0.044}{\rm(theo.)}$\\
      \end{tabular}
      \label{tab:results}
    \end{table}
  \end{minipage}
\end{center} 

\section{Upward-going muons and neutrino oscillations}

The probability of a neutrino produced as a $\nu_\mu$ remaining that way 
is :
\begin{equation}
  P(\nu_\mu \rightarrow \nu_\mu) = 1 - \sin^2(2\theta)
  \sin^2\left(\frac{1.27\Delta m^2 L}{E_\nu}\right),
  \label{eq:lovere}
\end{equation}
where $\Delta m^2$ is the square of the mass difference of the two
neutrino flavors, $\theta$ is the mixing angle, $L$ is the distance the
neutrino has traveled since production (the ``baseline''), and $E_\nu$
is the neutrino energy.  Of these, $L$ and $E_\nu$ are the experimental
handles on the problem.  While the $\nu_\mu$'s indirectly detected via
their daughter muons cannot be fully kinematically reconstructed in the
same way as the contained neutrino interactions, there are several ways
in which the upward-going muon signal is changed by oscillations.


First, $\nu_\mu \leftrightarrow \nu_\tau$ oscillations produce a lower
absolute flux of $\nu_\mu$.  The power of this test is reduced by the
large ($\sim 20\%$) theoretical uncertainty in the absolute flux, but
maximal mixing over many oscillation lengths would result in a 50\%
reduction in $\nu_\mu$ flux, so this test is still useful.


Second, as the zenith angle of the neutrino's arrival direction changes,
the baseline $L$ changes.  Note that only zenith angles from the nadir
($\cos\Theta=-1$) to the horizon ($\cos\Theta=0$) are accessible in the
upward-going muon analyses, as there is no way to tell if a
downward-going muon's parent is a neutrino or, far more likely, a cosmic
ray-induced muon.  Neutrinos arriving vertically upward travel roughly
13,000~km, while those coming from near the horizon originate only
$\sim$500~km away.  The effect of oscillations is more pronounced
for neutrinos traveling the longer distances (those having a larger
$L$ in Eq.~\ref{eq:lovere}), distorting the shape of the zenith angle
distribution in favor of fewer vertically upward-going
muons.  This shape comparison is fairly free of theoretical and
systematic errors, the bin-to-bin errors ranging from
$\pm(0.3-3.8)$\%\cite{skupthru}.  As can be seen in
Fig.~\ref{fig:fluxzen}b, the no-oscillation through-going muon predicted
shape does not fit the data well ($\chi^2/dof=18/9$), even after
allowing the overall normalization to float downwards by 10\%.


Finally, since the two available samples of upward-going muons have
parents of different energies, those having the smaller $E_\nu$ in
Eq.~\ref{eq:lovere} would have a larger chance to oscillate.  This
reduces the observed flux of stopping upward-going muons compared to
that of the higher energy upward through-going muons.  Forming a ratio
of the fluxes ${\mathcal{R}} = \Phi_s/\Phi_t$ is a useful way to make
this comparison, since much of the absolute theoretical uncertainty
divides out.  Thus, $\nu_\mu\leftrightarrow\nu_\tau$ oscillations reduce
$\Phi_s$, lowering $\mathcal{R}$.  As can be seen in
Table~\ref{tab:results}, the observed $\mathcal{R}$ is $2.9\sigma$ lower
than that expected for the no-oscillations case, and
Fig.~\ref{fig:fluxzen}a dramatically shows the loss of stopping muon
flux.

We estimated the most likely values of $\sin^22\theta$ and $\Delta m^2$
using using both the 10 zenith angle bins of upward through-going muon
flux and the 5 stopping upward-going muon zenith angle flux bins.  The
expected flux $(d\Phi/d\Omega)_{osc}$ for a given set of $\Delta
m^{2}$ and $\sin^{2}2\theta$ is calculated and the same binning is
applied to this flux as to the data.  To test the validity of a given
oscillation hypothesis, we minimize a $\chi^{2}$ which is defined as:
\begin{equation}
  \label{eq:chi2}
  \sum_{i=1}^{10}\left(\frac
  {\left(\frac{d\Phi_t}{d\Omega}\right)_{obs}^{i}-
    (1+\alpha_\mu)\left(\frac{d\Phi_t}{d\Omega}\right)_{osc}^{i}}
  {\sqrt{\sigma_{stat,i}^{2}+\sigma_{sys,i}^{2}}}\right)^{2}
  + \sum_{j=1}^{5}\left(\frac
  {\left(\frac{d\Phi_s}{d\Omega}\right)_{obs}^{j}-
    (1+\alpha_\mu)(1+\eta)\left(\frac{d\Phi_s}{d\Omega}\right)_{osc}^{j}}
  {\sqrt{\sigma_{stat,j}^{2}+\sigma_{sys,j}^{2}}}\right)^{2}
   + \left(\frac{\alpha_\mu}{\sigma_{\alpha_\mu}}\right)^{2}
   + \left(\frac{\eta}{\sigma_{\eta}}\right)^{2},
\end{equation}
\noindent
where $\sigma_{stat,i}$ ($\sigma_{sys,i}$) is the statistical
(experimental systematic) error in the observed flux
$(d\Phi/d\Omega)_{obs}^{i}$ for the $i$th bin, $(1+\alpha_\mu)$ is an
absolute normalization factor of the expected flux, and $(1 + \eta)$ is
a comparative normalization between stopping and through-going fluxes.
The absolute flux normalization error $\sigma_{\alpha_\mu}$ is estimated
to be $\pm$22~\%.  $\sigma_\eta$ is estimated to be
$^{+13}_{-12}$~\%, and is analogous to the systematic error in the
ratio $\mathcal{R}$.  $\sigma_{sys,i}$ ranges from $\pm(0.3-3.8)$\%.
Then, the minimum $\chi^{2} (\chi^{2}_{min})$ is sought on the
$\Delta m^{2}-\sin^{2}2\theta$ plane.

\begin{figure}[thbp]
  \centerline{
    \begin{minipage}[t]{3.5in}
      \epsfxsize 3.5 truein \epsfbox{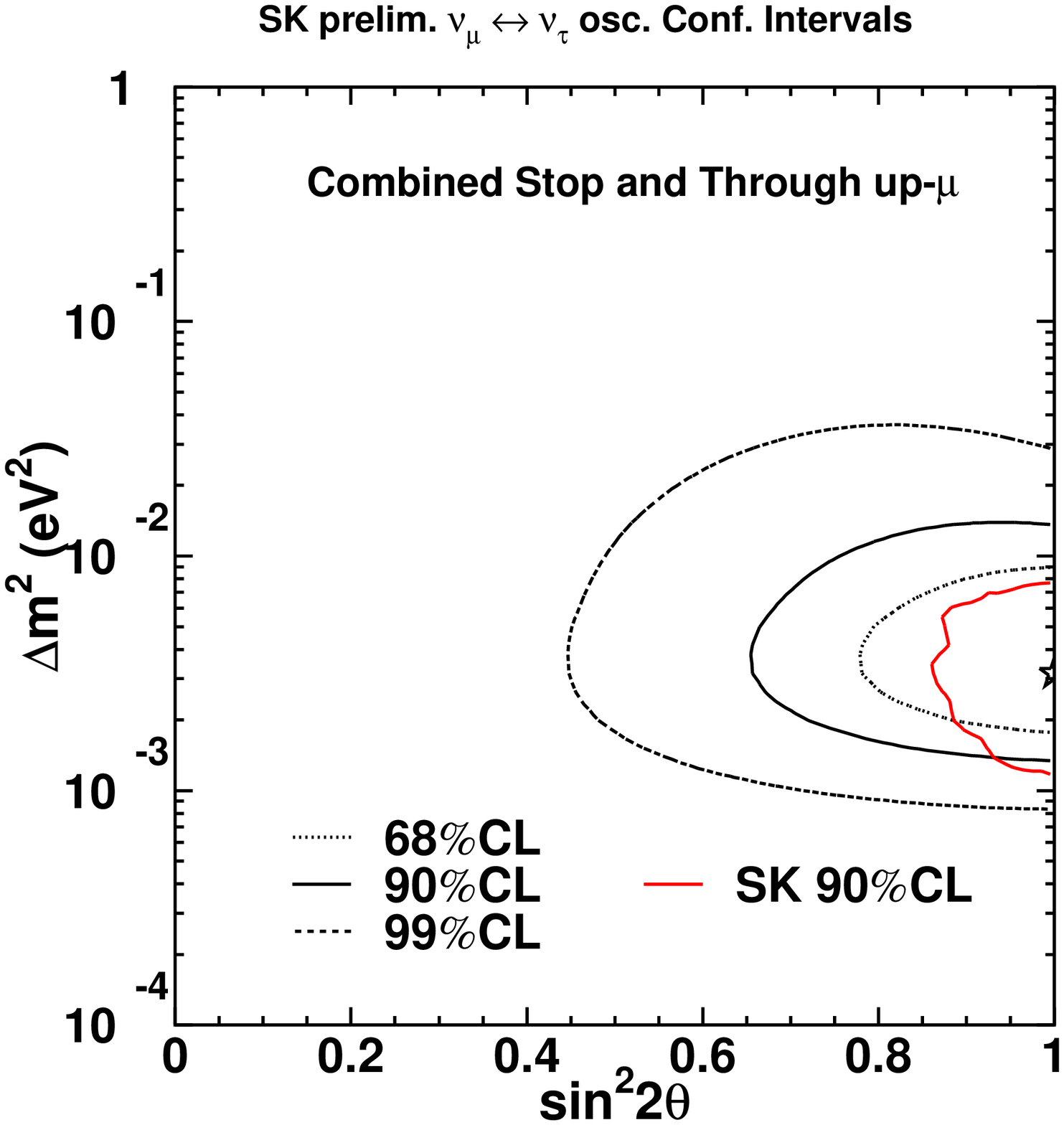}
    \end{minipage}
    \begin{minipage}[t]{3.5in}
      \epsfxsize 3.5 truein \epsfbox{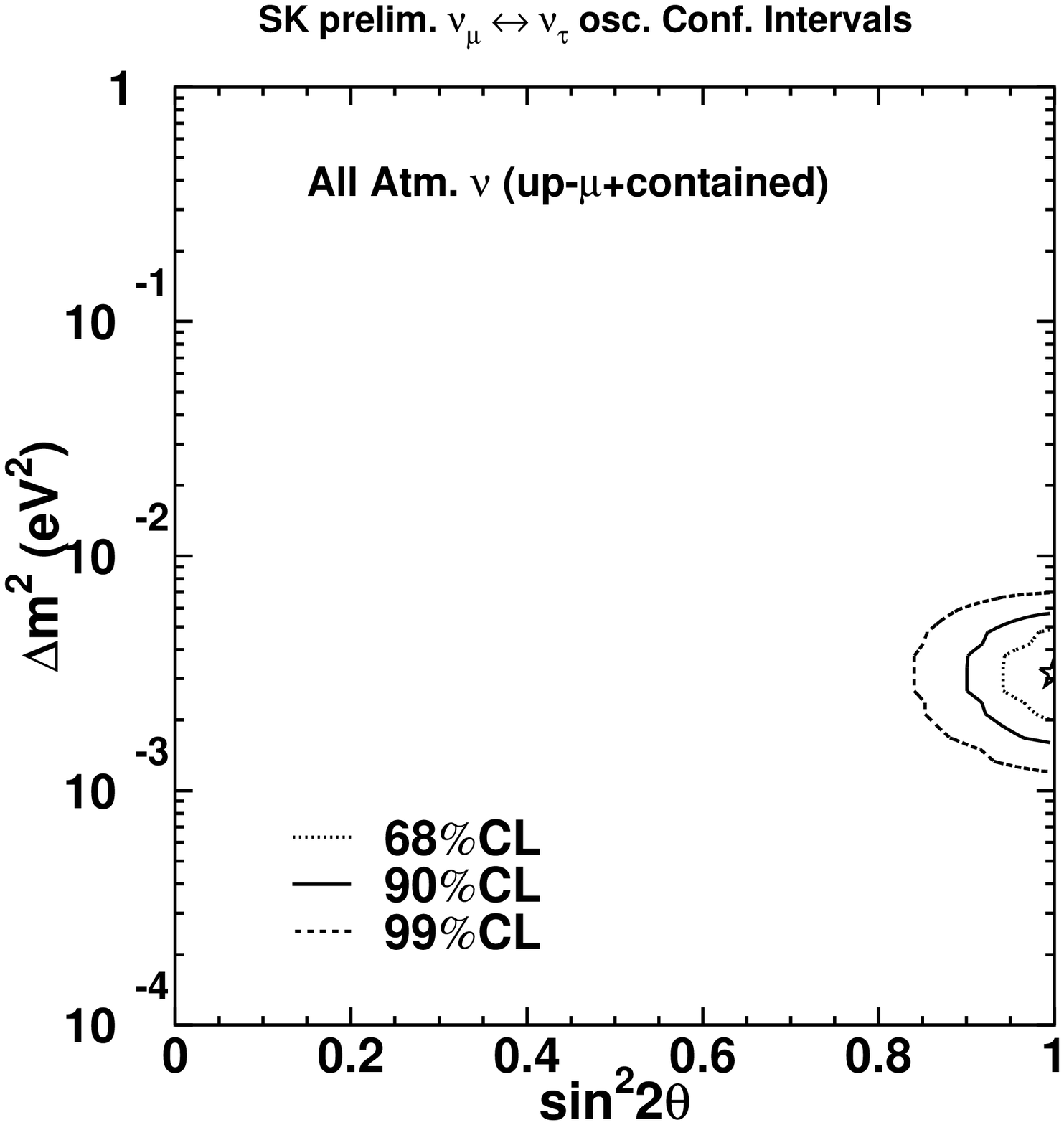}
    \end{minipage}
    }   
  \caption{a) the allowed region contours at 68\% (dotted contour), 
    90\% (thick solid), and 99\% (dashed) C.L. obtained by combining the
    Super-K upward through-going and stopping muon data in a fit on the
    ($\sin^{2}2\theta$,$\Delta{m}^{2}$) plane for the
    $\nu_{\mu}\leftrightarrow\nu_{\tau}$ oscillation hypothesis.  The
    star indicates the best fit point at $(\sin^2 2\theta, \Delta
    m^2)=(1.0, 3.2\times 10^{-3}{\rm{eV}^2})$.  Also shown is the
    allowed region contour (thin solid) at 90\% C.L. by the Super-K
    contained event analysis.  b) The same confidence
    intervals are presented for the global oscillation parameter fit of
    the Super-K contained data as well as the upward-going muons.  The
    star again represents the best fit point (which happens to be at the
    same point in parameter space as the upward-going muon fit alone).
    The allowed regions are to the right of the contours.}
  \label{fig:upmucontour}
\end{figure}

The no-oscillation case results in a $\chi^2/dof=41/15$, a poor
probability ($3.2\times10^{-4}$) of the null hypothesis.  However, the
best fit point of maximal mixing and $\Delta m^2=3.2 \times
10^{-3}$eV$^2$ matches the data well, as can be seen in
Fig.~\ref{fig:fluxzen}.  Fig.~\ref{fig:upmucontour}a shows the
confidence intervals in oscillation parameter space, taking into account
that the overall best fit is slightly outside the physical region at
$\sin^22\theta=1.1,\Delta m^2=3.7\times10^{-3}$eV$^2$.  These results
are consistent with those obtained in the analysis of Super-K's
contained atmospheric neutrino events\cite{skosc,messier-dpf}.  However,
they are obtained from an independent sample of much higher energy
atmospheric neutrinos.

Including both the upward-going muon terms from Eq.~\ref{eq:chi2} and
the contained event terms from \cite{skosc,messier-dpf} and properly
accounting for the joint systematic errors, a global oscillation fit
using all of the atmospheric neutrinos available to Super-K can then be
calculated.  This global $\chi^2$ is of the form :
\begin{equation}
  \label{eq:global}
  \chi^2 = \sum_{FC}^{65}~~+~~\sum_{PC\mu}^{5}~~+\sum_{upthru-\mu}^{10}+
           \sum_{upstop-\mu}^{5}+
           ~~\sum_{i=1}^{8}\left(\frac{\epsilon_i}{\sigma_{\epsilon i}}\right)^2,
\end{equation}
\noindent where $\epsilon_i/\sigma_{\epsilon i}$ are the systematic error
terms and their constraints.  In this fit, the $\alpha$ flux
normalization term is in common between the contained and upward-going
muon expected fluxes, although it is not constrained, following the
contained fit procedure.  The spectral index uncertainty ($E^{-\delta}$
from\cite{skosc}) is common and constrained.  The rest of the
stopping/through-going uncertainty from the $\eta$ term in
Eq.~\ref{eq:chi2} is placed in an $\eta_2$ term
($\sigma_{\eta2}=\pm7\%$), and a term $\eta_1$ ($\sigma_{\eta1}=\pm7\%)$
is introduced representing the systematic uncertainties between
contained and upward-muon analyses.  The no-oscillation case of
$\chi^2/dof=214.3/84$ strongly rules out the null hypothesis.
Minimizing the combined $\chi^2$ narrows the allowed region of parameter
space to close to maximal mixing and $1.5\times10^{-3} \lesssim \Delta
m^2 \lesssim 6\times10^{-3}$eV$^2$, as can be seen in
Fig.~\ref{fig:upmucontour}b.  This best fit point has
$\chi^2/dof=70.2/82$.  All the $\epsilon_i$ terms fall within one sigma of
the systematic error estimates.  The overall best fit point is
$\chi^2/dof=69.4/82$ outside the physical region at
$\sin^22\theta=1.05,\Delta m^2=3.2\times10^{-3}$eV$^2$, and the
confidence intervals have been appropriately corrected.

Making use of the higher energy $\nu_\mu$ interactions in the rock
surrounding Super-Kamiokande provides an independent and complementary
test of the $\nu_\mu\leftrightarrow\nu_\tau$ disappearance oscillations 
proposed to explain the atmospheric neutrino anomaly.  These 
upward-going muons confirm the conclusions reached in \cite{skosc}, and
a global fit to all the Super-K atmospheric neutrinos now available
provides a more precise measurement of the oscillation parameters
involved.


\begin{references}  

\bibitem{skosc}Y.~Fukuda {\it et al}, Phys. Rev. Lett. {\bf 81},
  1562 (1998).

\bibitem{messier-dpf}M.~Messier {\it et al}, these proceedings.

\bibitem{sksubgev}Y.~Fukuda {\it et al}, Phys. Lett. {\bf B433}, 9 (1998).

\bibitem{skmultigev}Y.~Fukuda {\it et al}, Phys. Lett. {\bf B436}, 33 (1998).

\bibitem{skupthru}Y.~Fukuda {\it et al}, Phys. Rev. Lett. {\bf 82}, 2644
  (1999).

\bibitem{up-pions}M.~Ambrosio {\it et al}, Astroparticle Physics
  {\bf 9} 105 (1998).

\bibitem{honda}M.~Honda {\it et al}, Phys. Rev. {\bf D52}, 4985
  (1995), Prog. Theor. Phys. Suppl. {\bf 123}, 483 (1996).

\bibitem{qe}C.H.~Llewellyn~Smith,  
Phys. Rep. {\bf 3}, 261 (1972).

\bibitem{sp}D.~Rein and L.M.~Seghal,   
  Ann. Phys. {\bf 133}, 79 (1981).

\bibitem{grv94}M.~Gl\H{u}ck, E.~Reya and A.~Vogt,
  Z. Phys. {\bf C67}, 433 (1995).

\bibitem{lohmann}W.~Lohmann, R.~Kopp and R.~Voss, CERN Yellow Report
  No. 85-03.

\bibitem{bartol}V.~Agrawal, T.K.~Gaisser, P.~Lipari, T.~Stanev,
  Phys. Rev. {\bf D53}, 1314 (1996).

\bibitem{cteq}J.~Botts {\it et al}, Phys. Lett. {\bf B304}, 159
  (1993); H.L.~Lai {\it et al}, Phys. Rev. {\bf D51}, 4763 (1995).

\bibitem{frati}W.~Frati {\it et al}, 
  Phys. Rev. {\bf D48}, 1140 (1993).

\end{references}
\end{document}